\documentclass[11pt,twoside]{article}
\usepackage{asp2010} 

\resetcounters 

\begin{document} 

\markboth{Cross et al.}{Notes on using 3D Extinction Maps}

\title{Extinction Maps in the WFAU
Archives} \author{Nicholas~Cross$^1$ and Mike~Read$^1$ 
\affil{$^1$Scottish Universities' Physics Alliance (SUPA), Institute for
Astronomy, School of Physics, University of Edinburgh, Royal Observatory, 
Blackford Hill, Edinburgh EH9 3HJ, UK}}


\begin{abstract} 
\end{abstract} 
  
\section{3D Extinction Maps} 

The VSA now includes 3D extinction maps, kindly given to us by Oscar Gonzalez,
based on \cite{Chen2013}, which is currently available through Vizier
\footnote{http://vizier.u-strasbg.fr/viz-bin/VizieR-3?-source=J/A\%2bA/550/A42}.
However, the maps in the VSA are a slightly higher resolution
($\Delta\,r=0.5$kpc, rather than 1kpc and $4\arcmin$ rather than $15\arcmin$)
and cover a slightly wider area, $-10<b<5$ and $-10<l<10$ thant the Vizier
versions, and can be used directly with the VVV survey data and other archived
data sets, and can therefore use the full power of the WFAU archive interfaces. 

The extinction maps consist of a 3-dimensional array of $E(J-K_s)$ and
$E(H-K_s)$ colour-excess terms, and their errors. These are derived from fitting a
model stellar population (using the Besancon model with thin disk, thick disk, bulge, and
spheroid) to the VVV and Glimpse data and using the necessary adjustment in
magnitude in the VVV J and Ks bands to match observations to derive an
extinction measure E(J-Ks). When J is not available (not enough sources bright 
enough, e.g. where the extinction is too great), the H and Ks bands were used
to derive E(H-Ks). Those derived from a wider wavelength range are more robust
and should be used in preference. 
 
Gonzalez suggests using the \citep{Nish2009} extinction law for $|b|<4$ and
the \citep{Card1989} extinction law for $|b|\geq4$. The $\frac{A_X}{E(J-K_s)}$
and $\frac{A_X}{E(H-K_s)}$ necessary to convert the colour-excesses to
extinction values in filter $X$ are calculated from these extinction terms.
Gonzalez provides the terms for the 2MASS filter
(http://mill.astro.puc.cl/BEAM/coffinfo.php) and we have derived the VISTA,
UKIRT-WFCAM, GLIMPSE, WISE and VST terms from there. 

We derive the Cardelli extinction terms for each filter using existing Cardelli
values from the Spanish Virtual Observatory Filter
Service\footnote{http://svo2.cab.inta-csic.es/theory/fps3/index.php?mode=browse}
\citep{SpanishVO}. The Spanish VO (and most other sources) give the correction
factor as $\frac{A_X}{E(B-V)}$. Using Eqn.~\ref{eq:cardconv} we are able to
convert these numbers to the required $\frac{A_X}{E(J-K_s)}$.

\begin{equation} 
\frac{A_X}{E(J-K_s)} = \frac{A_X}{E(B-V)}*\frac{E(B-V)}{A_Y}*\frac{A_Y}{E(J-Ks)}
\label{eq:cardconv}
\end{equation}

\noindent where $A_Y=A_{K_s}^{\rm 2MASS}$ for filters with $\lambda>1.6\mu\,m$
and $A_Y=A_{J}^{\rm 2MASS}$ for filters with $\lambda<1.6\mu\,m$. In each case
$\frac{E(B-V)}{A_Y}*\frac{A_Y}{E(J-Ks)}\sim1.9$. A similar conversion is done
for $\frac{A_X}{E(H-K_s)}$.

We make use of \cite{Nish2009} Table 1, to get the Nishiyama values for
UKIRT-WFCAM and VISTA $J$, $H$ and the Spitzer-IRAC values. For the VISTA $Ks$
and UKIRT-WFCAM $K$ we extrapolate using the 2MASS H and Ks values and the
effective wavelengths (from the Spanish VO). We also use the Spitzer-IRAC values
and effective wavelengths to calculate values for the WISE W1 and W2
filters, but we have not extrapolated the Nishiyama law outside
$1.2<\lambda<8.0\mu\,m$ for which the law was measured.

\section{WFAU Data Model}

We have designed a set of new tables to store the 3D extinction maps and
related data and to allow fast selection and use. 

\begin{itemize}
  \item \verb+ThreeDimExtinctionMaps+, a table of available maps.
  \item \verb+FilterExtinctionCoefficients+, a table of the coefficients to
  convert from colour-excess to extinction in each band.
  \item \verb+vvvBulgeExtMapCoords+, a table of the pixel coordinates for VVV
  bulge map.
  \item \verb+vvvBulge3DExtinctVals+, a table of the colour-excesses as a
  function of distance for VVV bulge map.
\end{itemize} 

The data is stored in a different way to Vizier. Instead of having a single
table, with a row for each position on the sky and every colour-excess at a
different distance having a separate column, we have split the data into two
tables: a table of pixels (\verb+vvvBulgeExtMapCoords+), which contains
positional information such as equatorial coordinates, Galactic coordinates,
Cartesian coordinates and Hierarchical Triangular Mesh Index and each pixel is
identified by a pixelID; a table of colour-excesses as a function of pixelID and
distance, r (in kpc). This design is more abstract than the Vizier design and 
can be used for other 3D extinction maps, such as recently created ones by
\citep[APOGEE;][]{APOGEE}, \citep[IPHAS;][]{IPHAS},
\citep[Gaia-ESO;][]{GaiaESOExt} and \citep[PanSTARRS;][]{PanSTARRS}. We will 
shortly be including the IPHAS extinction map.

The final table (\verb+FilterExtinctionCoefficients+) includes all the filter
coefficient terms. The table includes filters from 2MASS, VISTA-VIRCAM,
UKIRT-WFCAM, VST-OMEGACAM, Spitzer and WISE, covering most of the wide-field
surveys that have observed the region covered by this 3-D extinction map. For
each filter, there is an identifier, name, short-name, description and
wavelength range. There
are columns for $\frac{A_X}{E(J-K_s)}$, calculated in both the \cite{Nish2009} and
the \cite{Card1989} extinction laws: ($\bf aEJKsNish$) and ($\bf aEJKsCard$),
repectively, and the same for $\frac{A_X}{E(H-K_s)}$. Additionally, there is a column for $\frac{A_X}{E(B-V)}$ ($\bf aEBVCard$) taken
directly from the Spanish VO, which is used in the extrapolation for the
Cardelli law.

To match an object to the correct extinction map pixel, we have developed a new
table-valued function \verb+fGetPixelID(ra,dec)+, which finds the nearest pixel
within $6\arcmin$. The results are returned as a table, not a single scalar
value, which seems to be a more efficient way, possibly because SQL does the
rest of the selection first and then only calculates the tabular function on the
final set of rows, rather than calculating it on all positions early on.

\section{Using Dust Maps}

To match up the dust map to a small number of objects in the VVV or any other
overlapping survey, a CROSS APPLY, or OUTER APPLY can be used. These apply the
function to the data in the table defined before the apply term. CROSS APPLY is the equivalent of an INNER JOIN and OUTER
APPLY is the equivalent of an OUTER JOIN. 

\begin{verbatim}
SELECT s.sourceID, s.ksAperMag3, s.hmksPnt,c.pixelID
FROM vvvSource as s 
CROSS APPLY EXTINCT.dbo.fgetPixelID(ra,dec) as c
WHERE s.sourceID=515402469078 
\end{verbatim}
 
\begin{verbatim}
SELECT s.sourceID, s.ksAperMag3, s.hmksPnt,c.pixelID
FROM vvvSource as s 
OUTER APPLY EXTINCT.dbo.fgetPixelID(ra,dec) as c
WHERE s.sourceID=515402469078 
\end{verbatim}

\noindent where the \verb+EXTINCT.+ in front of the function is necessary
because the extinction tables and function are in a different database
(EXTINCT) from the VVVDR2 database and SQL needs to know where to find them.
Both the above cases return the same result:

\begin{verbatim}
sourceID	ksAperMag3	hmksPnt	pixelID
515402469078	+11.413226	+0.843723	30156
\end{verbatim}

However, sourceID 515793218971 lies in a region outside the table. Using CROSS
APPLY, returns no rows, whereas using an OUTER APPLY, returns a pixelID$=0$:

\begin{verbatim}
sourceID	ksAperMag3	hmksPnt	pixelID
515793218971	+18.182304	-9.999995E008	0
\end{verbatim}

To get the $J-Ks)$ colour excess for the first object at a particular distance,
e.g. 5kpc:

\begin{verbatim}
SELECT s.sourceID, s.ksAperMag3, s.hmksPnt,c.pixelID,e.ejks 
FROM vvvSource as s CROSS APPLY 
EXTINCT.dbo.fgetPixelID(ra,dec) as c,
EXTINCT..vvvBulge3DExtinctVals as e 
WHERE s.sourceID=515402469078 and e.pixelID=c.pixelID 
and e.r=0.5 
\end{verbatim}

\noindent returns, 

\begin{verbatim}
sourceID	ksAperMag3	hmksPnt	pixelID	ejks
515402469078	+11.413226	+0.843723	30156	+0.109000
\end{verbatim}

Equivalent queries can also be applied to other data sets that overlap the dust
maps, e.g. GLIMPSE, 

\begin{verbatim}
SELECT s.seqNo, s.mag1, s.mag2,c.pixelID,e.ejks 
FROM GLIMPSE..glimpse_hrc_inter as s 
CROSS APPLY EXTINCT.dbo.fgetPixelID(ra,dec) as c,
EXTINCT..vvvBulge3DExtinctVals as e 
WHERE s.seqNo=406 and e.pixelID=c.pixelID 
and e.r=0.5 
\end{verbatim}

\noindent and given that this object matches the previous VVV object, the
returned pixel and colour-excess are identical:

\begin{verbatim}
seqNo	mag1	mag2	pixelID	ejks
406	+10.550000	+10.563000	30156	+0.109000
\end{verbatim}

All of the above can be done for multiple objects in a single query and with
other selections applied too, e.g. selecting all sources in VVV with $0<l<3$ deg
and $-5<b<-4$ deg and $16<K_s<18$ mag and $(H-K_s)>2$ and finding the
colour-excess at $r=0.5$ kpc.

\begin{verbatim}
SELECT s.sourceID, s.ksAperMag3, s.hmksPnt,c.pixelID,e.ejks 
FROM vvvSource as s CROSS APPLY 
EXTINCT.dbo.fgetPixelID(ra,dec) as c,
EXTINCT..vvvBulge3DExtinctVals as e 
WHERE s.ksAperMag3>16. and s.ksAperMag3<18. and 
s.l>0. and s.l<3 and s.b>-5 and s.b<-4 and 
s.hmksPnt>2. and e.pixelID=c.pixelID 
and e.r=0.5 
\end{verbatim}

\noindent which returns 20 rows, which the user can try for themselves.

User created files can also be used, by selecting the {\it  enhanced version
of this form} in Freeform SQL and following the instructions. A query may look 
something like the following:

\begin{verbatim}
SELECT u.*,c.pixelID,e.ejks 
FROM #userTable as u CROSS APPLY 
EXTINCT.dbo.fgetPixelID(ra,dec) as c,
EXTINCT..vvvBulge3DExtinctVals as e 
WHERE e.pixelID=c.pixelID 
and e.r=0.5 
\end{verbatim}

However, for most scientific purposes, we need to calculate extinctions, not
just the colour-excesses. The following selection gives the extinction and
extinction corrected $K_s$ magnitude at a distance of $5$kpc.

\begin{verbatim}
SELECT s.sourceID, Ext.r, s.ksAperMag3, Ext.ejks, Ext.aKsCard, 
(s.ksAperMag3-Ext.aKsCard) as ksAperMag3ExtCor
FROM vvvSource as s 
CROSS APPLY EXTINCT.dbo.fgetPixelID(ra,dec) as c,
(SELECT e.*, (e.ejks*fks.aEJKsCard) as aKsCard
FROM EXTINCT..vvvBulge3DExtinctVals as e, 
EXTINCT..FilterExtinctionCoefficients as fks 
WHERE fks.filterID=9) as Ext
WHERE s.sourceID=515402469078 and Ext.pixelID=c.pixelID 
AND Ext.r=5.
\end{verbatim}

\noindent where it is better to calculate the extinction as derived sub-table,
where it can be used in multiple places. The colour-excess-to-extinction ratios
from the \verb+FilterExtinctionCoefficents+ table are used.

For VVVDR2 (and future releases), we have created a new table
\verb+vvvSourceExtinction+, which will contain matches between \verb+vvvSource+
and any extinction maps. This will contain the {\bf sourceID}, {\bf extMapID}
and {\bf extPixelID}, which will allow matches to multiple maps and so fast 
queries on hundreds of millions of sources, rather than tens of thousands will 
be possible. The function will still be useful for external databases where 
there are no matches to the VVV. The previous query can be done as:

\begin{verbatim}
SELECT s.sourceID, Ext.r, s.ksAperMag3, Ext.ejks, Ext.aKsCard, 
(s.ksAperMag3-Ext.aKsCard) as ksAperMag3ExtCor
FROM vvvSource as s, vvvSourceExtinction as c,
(SELECT e.*, (e.ejks*fks.aEJKsCard) as aKsCard
FROM EXTINCT..vvvBulge3DExtinctVals as e, 
EXTINCT..FilterExtinctionCoefficients as fks 
WHERE fks.filterID=9) as Ext
WHERE s.sourceID=515402469078 and s.sourceID=c.sourceID and
Ext.pixelID=c.extPixelID and c.extMapID=2 AND Ext.r=5.
\end{verbatim}

\subsection{SEDs}

\begin{verbatim}
SELECT s.sourceID, Ext.r, 
(s.zAperMag3-Ext.aZCard) as zAperMag3ExtCor,   
(s.yAperMag3-Ext.aYCard) as yAperMag3ExtCor, 
(s.jAperMag3-Ext.aJCard) as jAperMag3ExtCor,   
(s.hAperMag3-Ext.ahCard) as hAperMag3ExtCor, 
(s.ksAperMag3-Ext.aKsCard) as ksAperMag3ExtCor,   
(g.mag1-Ext.a34Card) as mag1ExtCor, 
(g.mag2-Ext.a45Card) as mag2ExtCor,   
(g.mag3-Ext.a58Card) as mag3ExtCor, 
(g.mag4-Ext.a80Card) as mag4ExtCor
FROM vvvSource as s 
CROSS APPLY EXTINCT.dbo.fgetPixelID(ra,dec) as c,
(SELECT e.*, (e.ejks*fz.aEJKsCard) as aZCard, 
(e.ejks*fy.aEJKsCard) as aYCard, (e.ejks*fj.aEJKsCard) as aJCard, 
(e.ejks*fh.aEJKsCard) as aHCard, (e.ejks*fks.aEJKsCard) as aKsCard, 
(e.ejks*f34.aEJKsCard) as a34Card, (e.ejks*f45.aEJKsCard) as a45Card, 
(e.ejks*f58.aEJKsCard) as a58Card, (e.ejks*f80.aEJKsCard) as a80Card
FROM EXTINCT..vvvBulge3DExtinctVals as e, 
EXTINCT..FilterExtinctionCoefficients as fz, 
EXTINCT..FilterExtinctionCoefficients as fy, 
EXTINCT..FilterExtinctionCoefficients as fj, 
EXTINCT..FilterExtinctionCoefficients as fh, 
EXTINCT..FilterExtinctionCoefficients as fks, 
EXTINCT..FilterExtinctionCoefficients as f34, 
EXTINCT..FilterExtinctionCoefficients as f45, 
EXTINCT..FilterExtinctionCoefficients as f58,
EXTINCT..FilterExtinctionCoefficients as f80 
WHERE fz.filterID=5 and fy.filterID=6 and fj.filterID=7 and 
fh.filterID=8 and fks.filterID=9 and f34.filterID=20 and 
f45.filterID=21 and f58.filterID=22 and f80.filterID=23) as Ext, 
GLIMPSE..glimpse_hrc_inter as g
WHERE s.sourceID=515402469078 and g.seqNo=406 and Ext.pixelID=c.pixelID 
AND Ext.r>5.
\end{verbatim}
 
\noindent where the corrected magnitude in each of the VISTA-VVV and
GLIMPSE (Spitser-IRAC) bands is calculated for each distance with 
$r>5$kpc. For full scientific usefulness the errors should also be calculated
and can use the errors in the colour-excesses.

\subsection{Colour-Magnitude Diagram}

Potentially even more useful is the locus of a star on the colour-magnitude
diagram. The following query can be used to select $M_J^e$, the extinction
corrected absolute J-band magnitude, and $(Z-Y)^e$, the extinction corrected
$(Z-Y)$ colour. 

\begin{verbatim}
SELECT Ext.rCor, 
(s.zAperMag3-Ext.aZCard-(s.yAperMag3-Ext.aYCard)) as zmyExtCor, 
(s.jAperMag3-Ext.aJCard-5.*log10(Ext.rCor)-10.) as absJExtCor
FROM vvvSource as s 
CROSS APPLY EXTINCT.dbo.fgetPixelID(ra,dec) as c,
(SELECT e.*, (e.ejks*fz.aEJKsCard) as aZCard, 
(e.ejks*fy.aEJKsCard) as aYCard, 
(e.ejks*fj.aEJKsCard) as aJCard,  
(e.r+0.005) as rCor
FROM EXTINCT..vvvBulge3DExtinctVals as e, 
EXTINCT..FilterExtinctionCoefficients as fz, 
EXTINCT..FilterExtinctionCoefficients as fy, 
EXTINCT..FilterExtinctionCoefficients as fj
WHERE fz.filterID=5 and fy.filterID=6 and 
fj.filterID=7) as Ext, 
GLIMPSE..glimpse_hrc_inter as g
WHERE s.sourceID=515402469078 and g.seqNo=406 and 
Ext.pixelID=c.pixelID  

\end{verbatim}

\noindent where we offset the distance $r$ by 5pc to give $rCor$. This offset
makes no difference to the extinction values, since it is much smaller than the
map resolution, but does allow us to get a value at $r\sim0$, without a
mathematical error.

\begin{figure}
\includegraphics[clip,width=60mm,angle=0]{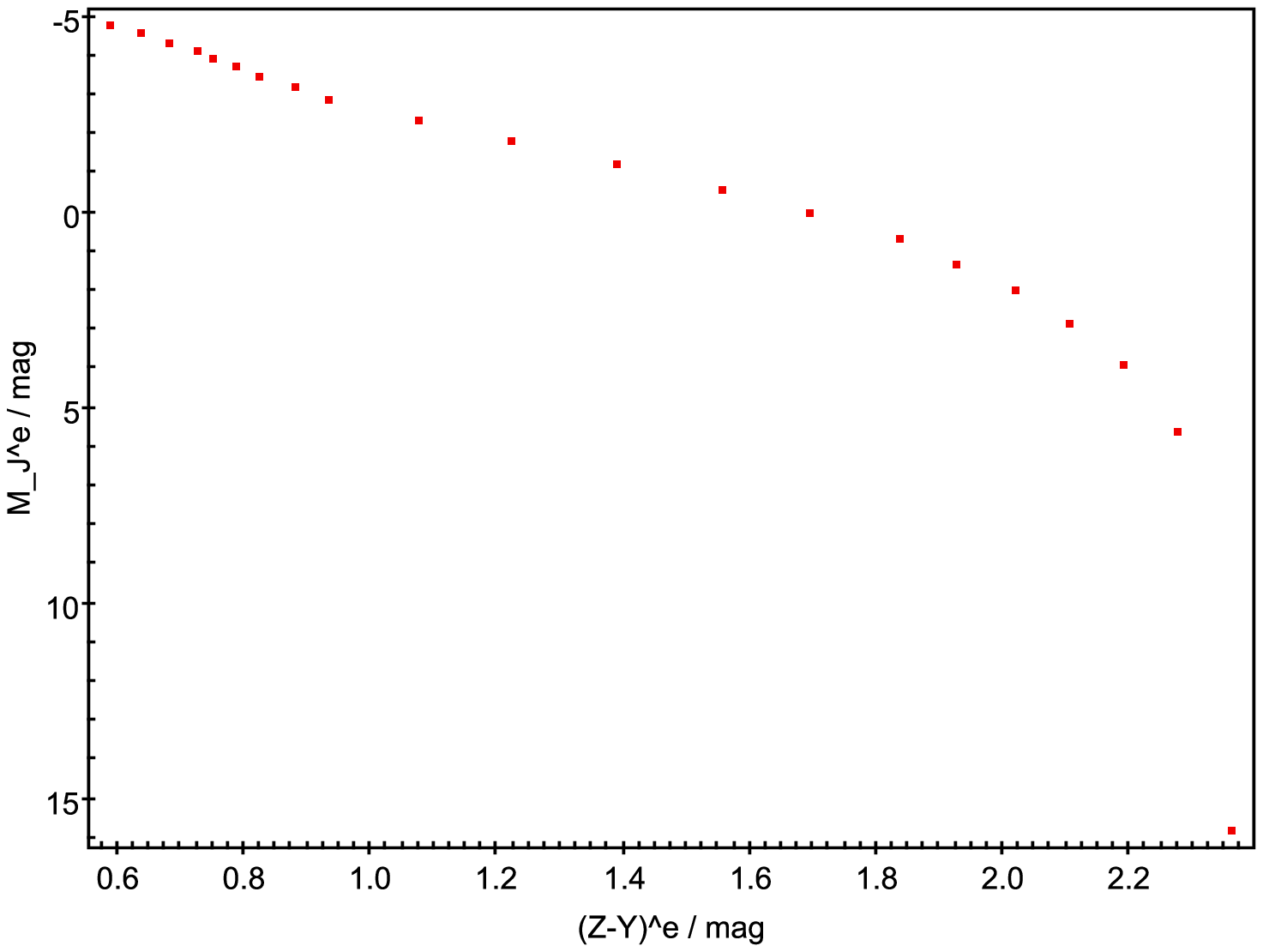}
\includegraphics[clip,width=60mm,angle=0]{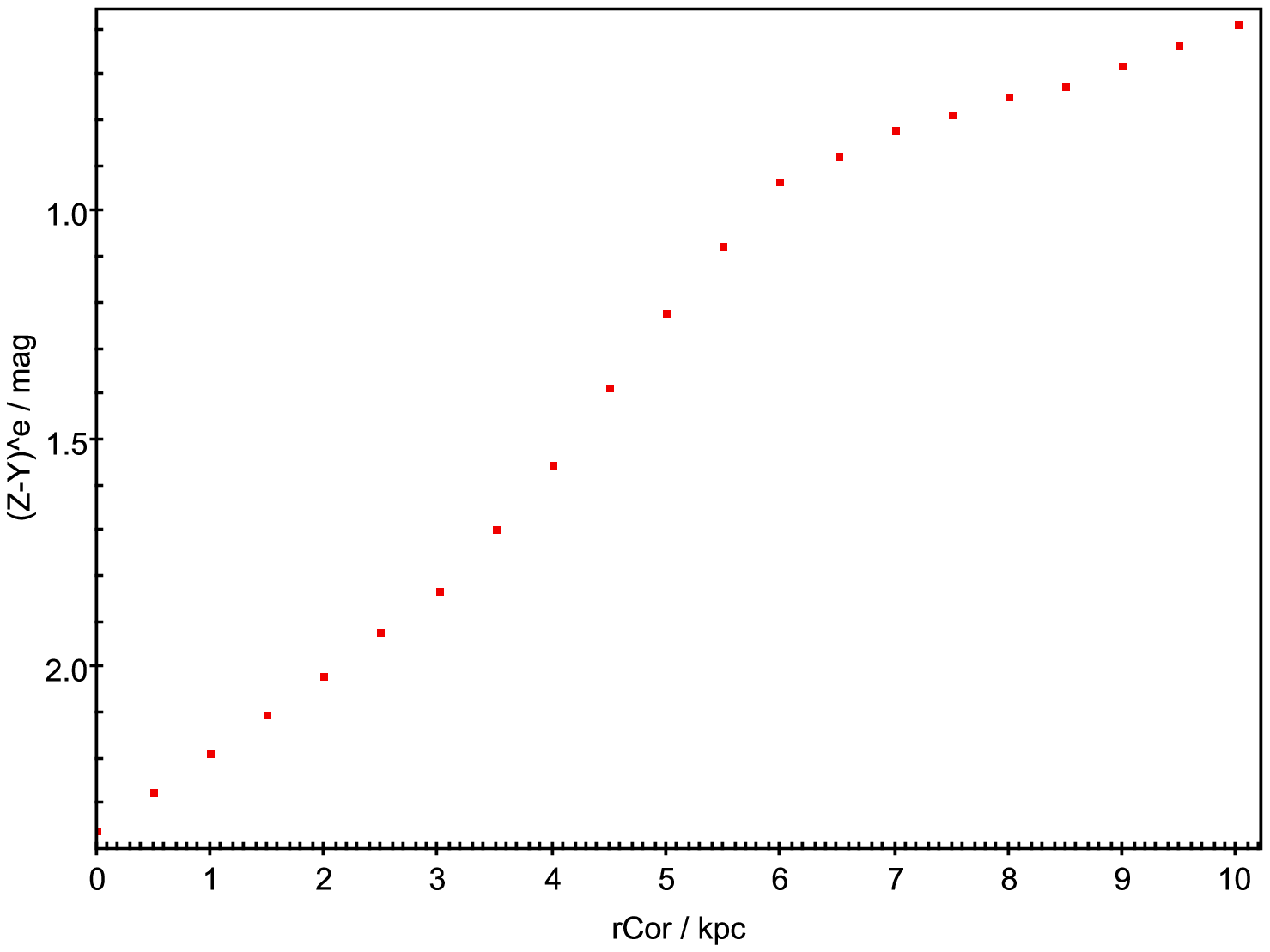}
  \caption{Left (a): Locus in the $(Z-Y)^e$ vs $M_J^e$ Colour-Magnitude Diagram
  of the VVV star with sourceID=515402469078. Right (b): $(Z-Y)^e$ colour as a
  function of distance (kpc).}
  \label{fig:CMD}
  \vspace{-5mm}
\end{figure}

 The locus of the star in the $(Z-Y)^e$ vs $M_J^e$ CMD is shown in
 Fig~\ref{fig:CMD}, as well as the variation in colour as a function of distance. 

\begin{figure}
\includegraphics[clip,width=100mm,angle=0]{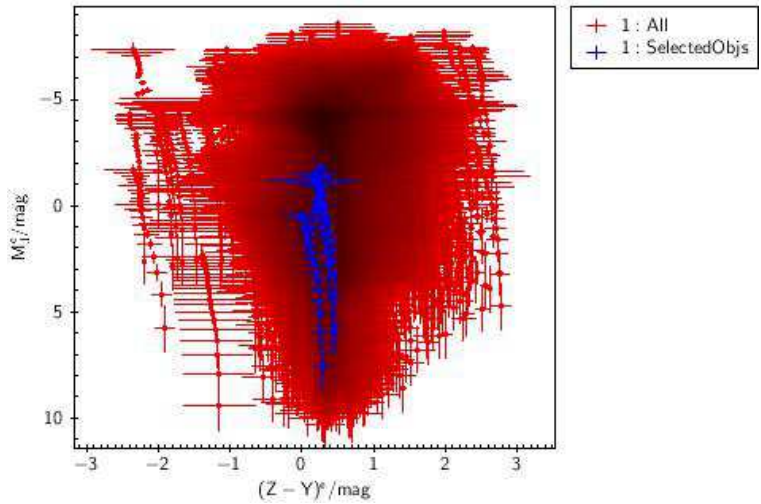}
  \caption{Locus in the $(Z-Y)^e$ vs $M_J^e$ Colour-Magnitude Diagram
  of $\sim10000$ VVV stars with errors included.}
  \label{fig:CMDuser}
  \vspace{-5mm}
\end{figure}

These queries can are very efficient and selections returning thousands of
objects can be returned within a few minutes. The \verb+vvvSource+ table has
$5\times10^8$ rows. The following is a selection using a set of $\sim12000$
objects already selected from \verb+vvvSource+ into a FITS file that has been
uploaded in the {\it enhanced query form}. The results are shown in
Fig.~\ref{fig:CMDuser}, including error-bars. A subsample of the data with
similar {\bf sourceID} values are highlighted in blue.

\begin{verbatim}
SELECT s.sourceID, corR, (jAperMag3-5.*log10(corR)-10.-aJ) as
absJ, (zAperMag3-aZ-(yAperMag3-aY)) as zmyPntCor, 
sqrt(jAperMag3Err*jAperMag3Err+aJErr*aJErr+(2.17*0.25/corR)* (2.17*0.25/corR))
as absJErr, sqrt(zAperMag3Err*zAperMag3Err+aZErr*aZErr+
yAperMag3Err*yAperMag3Err+aYErr*aYErr) as zmyPntCorErr 
FROM #usertable as u, vvvSourceExtinction as c, 
vvvSource as s, (select pixelID,(r+0.005) as corR,ejks,ejksErr,
(ejks*fz.aEJKsCard) as aZ, (ejksErr*fz.aEJKsCard) as aZErr, 
(ejks*fy.aEJKsCard) as aY, (ejksErr*fy.aEJKsCard) as aYErr, 
(ejks*fj.aEJKsCard) as aJ, (ejksErr*fj.aEJKsCard) as aJErr, 
(ejks*fh.aEJKsCard) as aH, (ejksErr*fh.aEJKsCard) as aHErr,
(ejks*fks.aEJKsCard) as aKs, (ejksErr*fks.aEJKsCard) as aKsErr 
from EXTINCT..vvvBulge3DExtinctVals as e, 
EXTINCT..FilterExtinctionCoefficients as fz, 
EXTINCT..FilterExtinctionCoefficients as fy, 
EXTINCT..FilterExtinctionCoefficients as fj, 
EXTINCT..FilterExtinctionCoefficients as fh, 
EXTINCT..FilterExtinctionCoefficients as fks where 
fz.filterID=5 and fy.filterID=6 and fj.filterID = 7 and 
fh.filterID=8 and fks.filterID=9) as Ext 
WHERE c.extPixelid=Ext.pixelid and c.sourceID=s.sourceID 
and c.extMapID=2 and u.sourceId=s.sourceID and yAperMag3>0. 
and jAperMag3>0. and Ext.corR>0.25 
\end{verbatim}

\bibliographystyle{asp2010}
\bibliography{ExtinctionNotes}

\end{document}